\documentclass[12pt]{article}
\usepackage{graphicx}
\usepackage{amsmath}
\input psfig.sty

\begin{document}

\begin{titlepage}

\title{\bf A Way to the Dark Side of the Universe through Extra
Dimensions \\ \vspace{0.2cm}}

\author{Je-An \ Gu\thanks{%
E-mail address: jagu@phys.ntu.edu.tw} \\
{\small Department of Physics, National Taiwan University, Taipei
106, Taiwan, R.O.C.}
\medskip
}

\date{\small \today}

\maketitle

\begin{abstract}
As indicated by Einstein's general relativity, matter and geometry
are two faces of a single nature. In our point of view, extra
dimensions, as a member of the {\em geometry face}, will be
treated as a part of the {\em matter face} when they are beyond
our poor vision, thereby providing dark energy sources
effectively. The geometrical structure and the evolution pattern
of extra dimensions therefore may play an important role in
cosmology. Various possible impacts of extra dimensions on
cosmology are investigated. In one way, the evolution of
homogeneous extra dimensions may contribute to dark energy,
driving the accelerating expansion of the universe. In the other
way, both the energy perturbations in the ordinary three-space,
combined with homogeneous extra dimensions, and the
inhomogeneities in the extra space may contribute to dark matter.
In this paper we wish to sketch the basic idea and show how extra
dimensions may lead to the dark side of our universe.
\end{abstract}


\end{titlepage}

\section{Introduction} \label{introduction}
It is strongly suggested by observational data that our universe
has the critical energy density and consists of 1/3 of dark matter
and 2/3 of dark energy (see e.g., Ref.~\cite{Turner:2002zb} and
references therein), where ``dark'' indicates the invisibility.
Even though it is generally not an elegant way to explain data via
something we cannot see, the avalanche of data, including those
from type Ia supernova measurements
\cite{Perlmutter:1999np,Riess:1998cb}, cosmic microwave
anisotropies \cite{Sievers}, galactic rotation curves, and surveys
of galaxies and clusters (providing the power spectrum of energy
density fluctuations), make it more and more convincing.
Nevertheless, we accordingly need to ask a question: {\em Why are
dark matter and dark energy so dark?}

This question reminds us another ``dark'' stuff, extra dimensions.
The existence of extra dimensions is required in various theories
beyond the standard model of particle physics, especially in the
theories for unifying gravity and other forces, such as
superstring theory. Extra dimensions should be ``hidden'' (or
``dark'') for consistency with observations. This common feature,
``invisible existence'', of dark energy, dark matter, and extra
dimensions provides us a hint that there may be some deep
relationship among them.

In this paper we show how extra dimensions may manifest themselves
as a source of energy in the ordinary three-space and lead to the
dark side of the universe. Basically homogeneous extra dimensions
will contribute to dark energy and may also provide some sort of
dark matter effectively if combined with the effects of
inhomogeneities in the ordinary three-space, and inhomogeneities
in the extra space will contribute to dark matter effectively. The
basic idea is sketched in the next section, and then we discuss in
Sec.\ \ref{homog ED to dark energy} how homogeneous extra
dimensions provide ``effective'' dark energy and influence the
evolution of the ordinary three-space, especially, producing the
accelerating expansion of the universe. The extra dimensions
employed throughout this paper are small and compact, as
introduced in the Kaluza-Klein theories.\footnote{Various
scenarios for hidden extra dimensions have been proposed, for
example, a brane world with large compact extra dimensions in
factorizable geometry proposed by Arkani-Hamed \emph{et al.}
\cite{Arkani-Hamed,Antoniadis:1990ew}, a brane world with extra
dimensions in warped nonfactorizable geometry proposed by Randall
and Sundrum \cite{Randall&Sundrum}, and small compact extra
dimensions in factorizable geometry as introduced in the
Kaluza-Klein theories \cite{Kaluza&Klein}.}

\section{A Sketch of the Idea} \label{idea sketch}
We consider a $(3+n+1)$-dimensional space-time where $n$ is the
number of extra spatial dimensions. The unperturbed metric tensor
$g_{\alpha \beta}$ ($\alpha , \beta =0,1,\ldots,3+n$), which
describes a universe with homogeneous, isotropic ordinary
three-space and extra space, is defined by
\begin{equation}
ds^2 = dt^2 - a^2(t) \left( \frac{dr_a^2}{1-k_a r_a^2} + r_a^2 d
\Omega_a^2 \right) - b^2(t) \left( \frac{dr_b^2}{1-k_b r_b^2} +
r_b^2 d \Omega_b^2 \right) , %
\label{unperturbed metric}
\end{equation}
where $a(t)$ and $b(t)$ are scale factors, and $k_a$ and $k_b$
relate to curvatures of the ordinary 3-space and the extra space,
respectively. The value of $r_b$ is set to be within the interval
$[0,1)$ for the compactness of extra dimensions. The perturbed
metric describing a lumpy universe is defined by
\begin{equation}
ds^2 = \left[ g_{\mu \nu} + \delta g_{\mu \nu}(x) \right] dx^{\mu}
dx^{\nu} + \left[ 1 + \delta_{b} (x^{\mu}) \right]^2 g_{pq} dx^p
dx^q , \label{perturbed metric}
\end{equation}
where $g_{\mu \nu}$ and $g_{pq}$ are unperturbed metric tensors,
while $\delta g_{\mu \nu}(x)$ and $\delta_b (x^{\mu})$
corresponding to perturbations, of the ordinary
$(3+1)$-dimensional space-time and the extra space, respectively.
As a convention, $x^{\mu ,\nu }$ and $x^{p,q}$ denote the
coordinates of the ordinary space-time and the extra space,
respectively, while $x$ denotes all the coordinates. For the sake
of simplicity the cross terms $dx^{\mu} dx^{p}$ are abandoned by
requiring the symmetry with respect to extra space inversion,
i.e., $x^{p} \rightarrow -x^{p}$. We note that the extra space is
kept to be homogeneous and isotropic after introducing
perturbations, so that we only need $\delta_b (x^{\mu})$, the
perturbation of the scale factor $b(t)$ as a function of the
coordinates of the ordinary space-time, to represent perturbations
of the extra space. On the contrary we have no symmetry
requirement for the perturbed ordinary space-time, and hence the
metric perturbations $\delta g_{\mu \nu}$ in general is a function
of all of the coordinates $\{x^{\gamma}, \gamma = 0,1,\ldots , 3+n
\}$.

Assuming that both the metric perturbations $\delta g_{\mu \nu}$
and $\delta_b$ are small, such that the Einstein equations from
the perturbed metric can be expanded with respect to these
perturbations, we obtain
\begin{eqnarray}
&& G_{\alpha \beta} = 8 \pi \bar{G} T_{\alpha \beta} = 8 \pi \bar{G} %
   \left[ T_{\alpha \beta}^{(0)}(t) + \delta T_{\alpha \beta} %
   \left( x \right) \right] \label{expanded Eintein eq} \\
&& = G_{\alpha \beta}^{(0)} \left[ g_{\mu \nu}(t) \right] + %
     G_{\alpha \beta}^{(1)} \left[ g_{\mu \nu}(t),\delta g_{\mu \nu}
                            \left( x \right) \right] + %
     G_{\alpha \beta}^{(2)} \left[ g_{\mu \nu}(t) , b(t) \right] \\ %
&& \quad
   + G_{\alpha \beta}^{(3)} \left[ g_{\mu \nu}(t) , \delta g_{\mu \nu} %
                            \left( x \right) , b(t) \right] + %
     G_{\alpha \beta}^{(4)} \left[ g_{\mu \nu}(t),\delta g_{\mu \nu}
                            \left( x \right), b(t) ,
                            \delta _b \left( x^{\mu} \right) \right] , \nonumber %
\end{eqnarray}
where $\bar{G}$ is the gravitational constant in the
higher-dimensional space-time, and $T_{\alpha \beta}$ denotes the
energy-momentum tensor, $T_{\alpha \beta}^{(0)}(t)$ the
unperturbed, and $\delta T_{\alpha \beta}(x)$ the perturbed one.
The first two terms in the above expansion of the Einstein tensor,
$G_{\alpha \beta}^{(0)}$ and $G_{\alpha \beta}^{(1)}$, are exactly
the unperturbed and the perturbed Einstein tensor, respectively,
of the ordinary $(3+1)$-dimensional space-time. In contrast, 
$G_{\alpha \beta}^{(2)}$, $G_{\alpha \beta}^{(3)}$, and $G_{\alpha
\beta}^{(4)}$ are additional terms coming from extra dimensions.
In our point of view, if observers are too blind to see extra
dimensions, these three additional terms will be automatically
moved to the right-hand side of the Einstein equations
(\ref{expanded Eintein eq}) and treated as some sort of energy
source, thereby contributing an ``effective'' part to the
energy-momentum tensor. In particular, $G_{\alpha \beta}^{(2)}$ is
smoothly distributed in the space and hence contributes to dark
energy, while $G_{\alpha \beta}^{(3)}$ and $G_{\alpha
\beta}^{(4)}$ have the spatial dependence and contribute to dark
matter. In the above discussion we have sketched the main idea. As
a demonstration of this idea, we will show in the next section how
homogeneous extra dimensions can lead to ``effective'' dark energy
and consequently change the evolution pattern of a
(nonrelativistic-) matter-dominated universe.\footnote{The part of
``effective'' dark matter originated from extra dimensions is
currently under investigation, and will not be discussed in detail
in the rest of this paper.}

\section{Dark Energy from Homogeneous Extra Dimensions} %
\label{homog ED to dark energy}
We consider in this section the case of a (3+n+1)-dimensional
space-time described by the unperturbed metric defined in Eq.\
(\ref{unperturbed metric}), i.e., both the ordinary three-space
and the extra space are homogeneous and isotropic. Assuming that
the matter content in this higher-dimensional space is a perfect
fluid with the energy-momentum tensor
\begin{equation}
T^{\alpha}_{\; \; \beta}=diag(\bar{\rho},-\bar{p}_a, \ldots ,
-\bar{p}_b , \ldots ) ,
\end{equation}
we can write the Einstein equations as
\begin{equation}
3 \left[ \left( \frac{\dot{a}}{a}\right)^2 + \frac{k_a}{a^2}
\right] + \frac{n(n-1)}{2} \left[ \left(
\frac{\dot{b}}{b}\right)^2 + \frac{k_b}{b^2} \right] +
3n\frac{\dot{a}}{a} \frac{\dot{b}}{b} = 8 \pi \bar{G} \bar{\rho}
\, , \label{G00 eq with homog ED}
\end{equation}
\begin{eqnarray}
2 \frac{\ddot{a}}{a} + n\frac{\ddot{b}}{b} + \left[ \left(
\frac{\dot{a}}{a} \right)^2 + \frac{k_a}{a^2} \right] +
\frac{n(n-1)}{2} \left[ \left( \frac{\dot{b}}{b}\right)^2 +
\frac{k_b}{b^2} \right] && \nonumber \\
+ 2n \left( \frac{\dot{a}}{a} \right) \left( \frac{\dot{b}}{b}
\right) = - 8 \pi \bar{G} \bar{p}_a \, , && %
\label{Gii eq with homog ED}
\end{eqnarray}
\begin{eqnarray}
3 \frac{\ddot{a}}{a} + (n-1)\frac{\ddot{b}}{b} + 3 \left[ \left(
\frac{\dot{a}}{a} \right)^2 + \frac{k_a}{a^2} \right] +
\frac{(n-1)(n-2)}{2} \left[ \left( \frac{\dot{b}}{b}\right)^2 +
\frac{k_b}{b^2} \right] && \nonumber \\
+ 3(n-1)\left( \frac{\dot{a}}{a} \right)
\left( \frac{\dot{b}}{b} \right) = - 8 \pi \bar{G} \bar{p}_b \, , && %
\label{Gjj eq with homog ED}
\end{eqnarray}
where $\bar{\rho}$ is and the energy density in the
higher-dimensional world, and $\bar{p}_a$ and $\bar{p}_b$ are the
pressures in the ordinary three-space and the extra space,
respectively.

In the previous work by Gu and Hwang \cite{Gu:2001ni}, the case
with $k_a=k_b=0$ was considered, in which the accelerating
expansion of the present (nonrelativistic-) matter-dominated
universe was proposed to be generated along with the evolution of
extra dimensions. Here we also focus on a matter-dominated
universe, setting $\bar{p}_a$ and $\bar{p}_b$ to zero accordingly,
but consider a more general case in which only $k_b=0$ is assumed
while $k_a$ is treated as a free parameter. %
In this case Eqs.\ (\ref{Gii eq with homog ED}) and (\ref{Gjj eq
with homog ED}) can be rearranged to become
\begin{equation}
(n+2) \frac{\ddot{a}}{a} + (2n+1) \left[ \left( \frac{\dot{a}}{a}
\right)^2 + \frac{k_a}{a^2} \right] +
n(n-1)\frac{\dot{a}}{a}\frac{\dot{b}}{b} - \frac{n(n-1)}{2} \left(
\frac{\dot{b}}{b} \right)^2 = 0 , \, %
\label{ddot-a}
\end{equation}
\begin{equation}
(n+2) \frac{\ddot{b}}{b} - 3 \left[ \left( \frac{\dot{a}}{a}
\right)^2 + \frac{k_a}{a^2} \right] + 6
\frac{\dot{a}}{a}\frac{\dot{b}}{b} + \frac{(n-1)(n+4)}{2} \left(
\frac{\dot{b}}{b} \right)^2 = 0 \, , %
\label{ddot-b}
\end{equation}
and then we can rewrite the Einstein equations, using new
variables $u(t) \equiv \dot{a}/a$ and $v(t) \equiv \dot{b}/b$, as
\begin{equation}
3 \left( u^2 + \frac{k_a}{a^2} \right) + 3nuv +
\frac{n(n-1)}{2}v^2 = 8 \pi \bar{G} \bar{\rho} \, , %
\label{G00 eq with u v}
\end{equation}
\begin{equation}
(n+2)\dot{u} + 3(n+1)u^2 + (2n+1)\frac{k_a}{a^2} + n(n-1)uv
-\frac{n(n-1)}{2}v^2 = 0 \, , \label{dot-u}
\end{equation}
\begin{equation}
(n+2)\dot{v} - 3\left( u^2 +\frac{k_a}{a^2} \right) + 6uv +
\frac{n(n+5)}{2}v^2 = 0 \label{dot-v} \, .
\end{equation}

Before getting numerical solutions, we use simple analytical
operations to extract, from the above Einstein equations,
essential features of these equations and the evolution patterns
governed by them. We first obtain, from Eq.\ (\ref{ddot-a}),
conditions for the accelerating and the decelerating expansion:
\begin{eqnarray}
         && > 0 \quad , \quad v/u < J_{-} \nonumber \\
\ddot{a} && < 0 \quad , \quad J_{-} < v/u < J_{+} \; , %
                \label{ddot a condition} \\
         && > 0 \quad , \quad v/u > J_{+} \nonumber %
\end{eqnarray}
where
\begin{equation}
J_{\pm} \equiv 1 \pm  \sqrt{\frac{(n+1)(n+2) + 
2(2n+1) k_a / \left( a^2 u^2 \right) }{n(n-1)}} \, . %
\label{J definition}
\end{equation}
We then read off from Eq.\ (\ref{G00 eq with u v}) that the
condition for positive energy density $\bar{\rho}$ is
\begin{equation}
v/u > K_{+} \quad \mbox{or} \quad v/u < K_{-} \, , %
\label{positive rho condition}
\end{equation}
where
\begin{equation}
K_{\pm} \equiv -\frac{3}{n-1} \pm \sqrt{\frac{3}{n(n-1)} \left(
\frac{n+2}{n-1} - \frac{2k_a}{a^2 u^2} \right)} \; . %
\label{K definition}
\end{equation}

Observing Eqs.\ (\ref{ddot a condition})--(\ref{K definition}), we
notice that variables $v/u$ and $k_a / (a^2 u^2)$ play essential
roles in the above expressions of these conditions. These two
essential variables can also be recognized from Eqs. (\ref{G00 eq
with u v})--(\ref{dot-v}), which tell us that values of all the
quantities in them are determined, up to an overall factor related
to the initial value of $u$, once the values of $v/u$ and $k_a /
(a^2 u^2)$ are given. It is therefore a good way to analyze the
evolution of the universe governed by Eqs.\ (\ref{G00 eq with u
v})--(\ref{dot-v}) via a two-dimensional diagram described by
$v/u$ and $k_a / (a^2 u^2)$.

\begin{figure}[h!]
\centerline{\psfig{figure=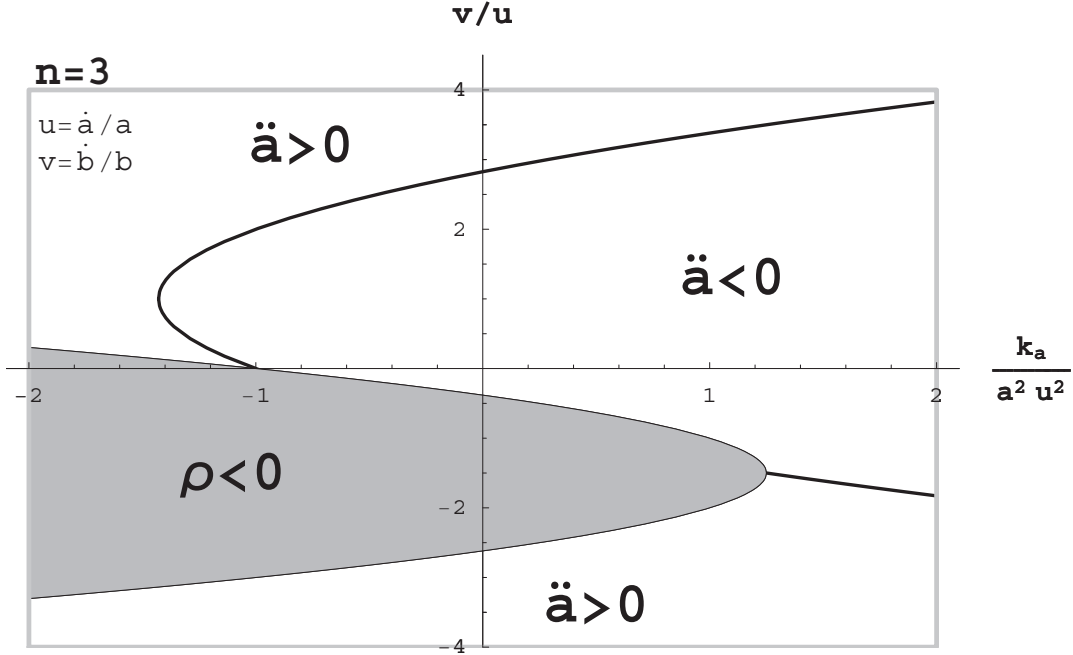,height=3.45in}}
\caption{Conditions for various signs of energy density $\rho$ and
acceleration $\ddot{a}$ are illustrated, where the number of extra
dimensions $n$ is specified to be three.} %
\label{conditions plot} %
\end{figure}

\begin{figure}[t!]
\centerline{\psfig{figure=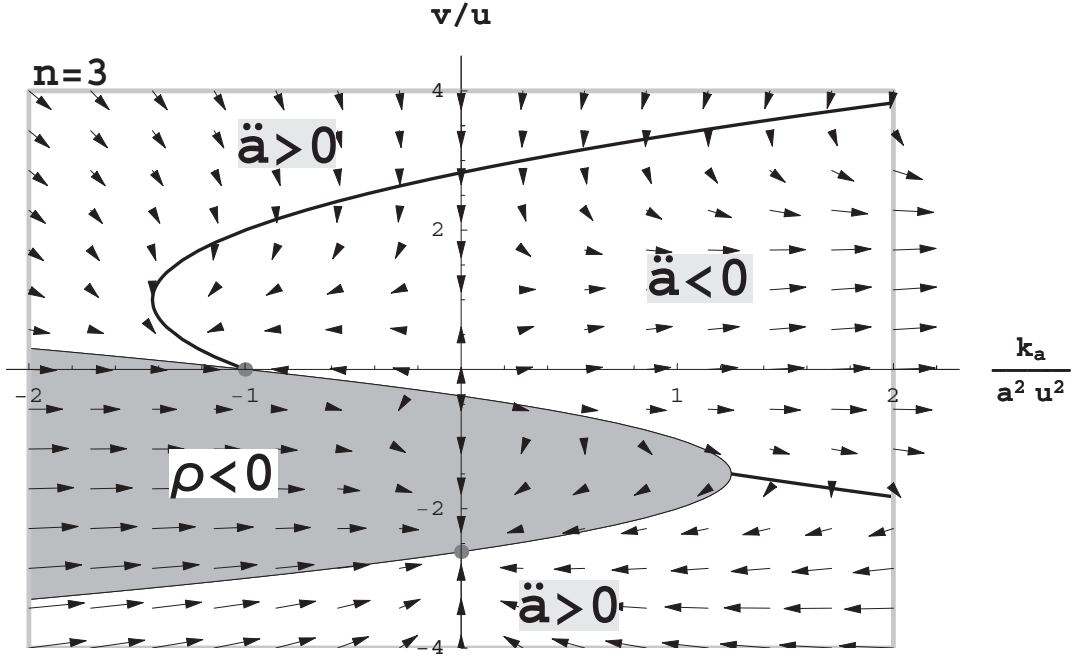,height=3.45in}} %
\caption{Flow vectors in $(k_a/(a^2u^2) , v/u)$-diagram are
plotted. Two grey dots denote two ``attractors'' at (-1,0) and
$(0,-\left[ 3+\sqrt{3(n+2)/n}\right] / (n-1))$ (where $n=3$), respectively.} %
\label{flow plot} %
\end{figure}



Conditions in Eqs.\ (\ref{ddot a condition})--(\ref{K definition})
are summarized in Fig.\ \ref{conditions plot}, where the number of
extra dimensions $n$ is specified to be three as an example. The
grey area denoted by ``$\rho <0$'' is a forbidden region if
positive energy density is required. In addition, flow vectors in
$(k_a/(a^2u^2) , v/u)$-diagram, as determined by Eqs.\
(\ref{dot-u}) and (\ref{dot-v}), are plotted in Fig.\ \ref{flow
plot} (where $n=3$). There are two ``attractors'' denoted by grey
dots in the flow diagram: one at $(k_a/(a^2u^2) , v/u)=(-1,0)$,
and the other at $(k_a/(a^2u^2) , v/u)=(0,-\left[
3+\sqrt{3(n+2)/n}\right] / (n-1))$.\footnote{Attractors are stable
fixed points toward which the nearby points (or ``state'') tend to
flow.} The attractor at $(-1,0)$ is on the margin of the forbidden
region (i.e., indicating $\rho =0$) and corresponding to a state
of the higher-dimensional universe entailing stable extra
dimensions and vanishing $\ddot{a}$. We note that the existence of
solutions corresponding to stable extra dimensions is a good
feature for building models in a higher-dimensional space-time.
The other attractor is also on the margin, with zero energy
density, of the forbidden region, entailing collapsing extra
dimensions and positive acceleration.

\begin{figure}[t]
\centerline{\psfig{figure=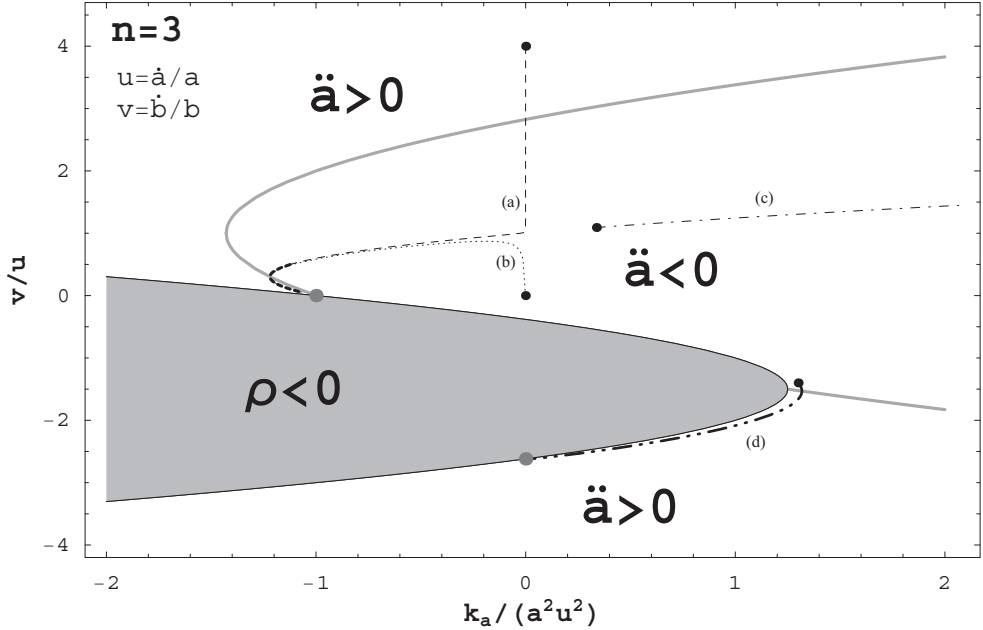,height=3.3in}} %
\caption{Four trajectories corresponding to four numerical
solutions with respect to initial conditions,
$(k_a/(a^2u^2),v/u)=$ (a) $(-0.0001,4)$, (b) $(-0.001,0)$, (c)
$(0.0001,4)$, and (d) $(1.3,-1.4)$, are plotted, where the black
dot at one end of each trajectory denotes the initial position.
(As in Fig.\ \ref{flow plot}, two grey dots are ``attractors'' and
$n=3$.)} %
\label{trajectory plot} %
\end{figure}

For a concrete illustration, we now solve Eqs.\ (\ref{G00 eq with
u v})--(\ref{dot-v}) numerically for the case of $n=3$. We plot in
Fig.\ \ref{trajectory plot} four trajectories corresponding to
four numerical solutions with respect to initial conditions,
$(k_a/(a^2u^2),v/u)=$ (a) $(-0.0001,4)$, (b) $(-0.001,0)$, (c)
$(0.0001,4)$, and (d) $(1.3,-1.4)$. These four trajectories
represent four different kinds of evolution path: %
\renewcommand{\theenumi}{\alph{enumi}}
\begin{enumerate}
\item {\bf acceleration $\rightarrow$ deceleration $\rightarrow$ acceleration},
      eventually approaching the attractor at $(-1,0)$ with stable extra
      dimensions and zero acceleration, possessing negative spatial curvature.
\item {\bf deceleration $\rightarrow$ acceleration}, eventually merging to
      the trajectory (a) and approaching the attractor at $(-1,0)$ with stable
      extra dimensions and zero acceleration, possessing negative spatial curvature.
\item {\bf eternal deceleration}, possessing increasing positive curvature contribution.
\item {\bf deceleration $\rightarrow$ acceleration}, eventually approaching
      the attractor at \\
      $(0,-\left[ 3+\sqrt{3(n+2)/n}\right] / (n-1))$ with collapsing
      extra dimensions, possessing decreasing positive curvature contribution.
\end{enumerate} %
\renewcommand{\theenumi}{\arabic{enumi}}
It is therefore indicated that there are many possibilities of
evolution patterns in this higher-dimensional universe, in
contrast to the unique manner of evolution, eternally decelerating
expansion, for a matter-dominated universe in the standard
cosmology without extra dimensions.

\section{Discussion and summary} \label{summary}
In this paper we make a point that there may be a deep
relationship between \mbox{``hidden''} (or ``dark'') extra
dimensions and the dark side of the universe, i.e., dark matter
and dark energy. This conjecture is based on Einstein's general
relativity, which indicates an important aspect that matter (with
energy and momentum) and geometrical structures of a space-time
are two faces of a single nature, to be called {\em matter face}
and {\em geometry face}, respectively. In our point of view, if
there exists a part of the {\em geometry face} which is beyond our
poor vision, this missing part will be treated as a member of the
{\em matter face}, and consequently provide mysterious, dark,
``effective'' energy sources. A possible missing part of the {\em
geometry face} we consider in this paper is the existence of extra
dimensions. This idea is sketched in Sec.~\ref{idea sketch} via
analyzing the Einstein equations, including perturbations of both
the metric tensor and the energy-momentum tensor, for a
higher-dimensional world. We conclude that extra dimensions may
manifest themselves as a source of energy in the ordinary
three-space, such as ``effective'' dark energy, under the
consideration of homogeneous extra dimensions, and ``effective''
dark matter, as contributed by inhomogeneities in the extra space
or the ordinary three-space.

As a particular demonstration of the general idea, we consider in
Sec.~\ref{homog ED to dark energy} a (non-relativistic-)
matter-dominated universe with homogeneous extra dimensions and
show that the evolution of homogeneous extra dimensions can lead
to ``effective'' dark energy and consequently change the evolution
pattern of the universe. There are many possibilities of evolution
patterns in this higher-dimensional universe, in contrast to the
unique way of evolution, eternally decelerating expansion, for a
matter-dominated universe in the standard cosmology without extra
dimensions. It needs further detailed studies to determine which
evolution pattern can appropriately describe our universe. In
addition, there are various possible realizations of this idea
worthy of further quests, and some are currently under our
investigation.

As mentioned in Sec.\ \ref{introduction}, this work is motivated
by a fundamental question: {\em Why are dark matter and dark
energy so dark?} Through the preliminary studies of the general
idea discussed in this paper, here comes up a possible answer:
{\em Dark matter and dark energy are generated from the extra
dimensions, a nature of geometry we are too blind to see.} This
simple answer indicates an intriguing possibility of unifying
these two kinds of dark entities, extra dimensions and dark energy
sources, into one.


\newpage

\section*{Acknowledgements}
The author wishes to thank Professor W-Y.~P.~Hwang for helpful
discussions. This work was supported by Taiwan CosPA project of
the Ministry of Education (MOE 91-N-FA01-1-4-0).


\end{document}